\documentclass[]{emulateapj} 
\usepackage{latexsym}
\usepackage{mathrsfs}
\usepackage{times} 
\usepackage{graphics}
\usepackage{graphics}

\newcommand{\scrW}{\mathcal{W}}

\newcommand{\scrR}{\mathcal{R}}
\newcommand{\scrv}{\mathcal{V}}

\newcommand{\sci}{Science}

\begin{document}
\shorttitle{Microquasars as head-tails sources} \title{Blazing trails:
Microquasars as head-tail sources and the seeding of magnetized
plasma into the ISM}

\shortauthors{Heinz, Grimm, Sunyaev, \& Fender}
\author{{Heinz}, S.}
\affil{University of Wisconsin-Madison}
\affil{Department of Astronomy, 6508 Sterling Hall, 475 N.~Charter St., Madison, WI 53593}
\email{heinzs@astro.wisc.edu}
\author{Grimm, H.J.}
\affil{Harvard Smithsonian Center for Astrophysics}
\affil{60 Garden St., Cambridge, MA 02138}
\author{Sunyaev, R.A.}
\affil{Max-Planck-Institute for Astrophysics}
\affil{Karl-Schwarzschil-Str. 1, 85741 Garching, Germany}
\author{Fender, R.P.}
\affil{University of Shouthampton}
\affil{Department of Astronomy, Building B46, Southampton, Hampshire
  SO17 1BJ, United Kingdom}

\begin{abstract}
We discuss the dynamics of microquasar jets in the interstellar
medium, with specific focus on the effects of the X-ray binaries'
space velocity with respect to the local Galactic standard of rest. We
argue that, during late stages in the evolution of large scale radio
nebulae around microquasars, the ram pressure of the interstellar
medium due to the microquasar's space velocity becomes important and
that microquasars with high velocities form the Galactic equivalent of
extragalactic head--tail sources, i.e., that they leave behind trails
of stripped radio plasma.  Because of their higher space velocities,
low--mass X-ray binaries are more likely to leave trails than
high--mass X--ray binaries.  We show that the volume of radio plasma
released by microquasars over the history of the Galaxy is comparable
to the disk volume and argue that a fraction of a few percent of the
radio plasma left behind by the X-ray binary is likely mixed with the
neutral phases of the ISM before the plasma is removed from the disk
by buoyancy.  Because the formation of microquasars is an unavoidable
by-product of star formation, and because they can travel far from
their birth places, their activity likely has important consequences
for the evolution of magnetic fields in forming galaxies.  We show
that radio emission from the plasma inside the trail should be
detectable at low frequencies.  We suggest that LMXBs with high
detected proper motions like XTE J1118+480 will be the best candidates
for such a search.
\end{abstract}

\section{Introduction}
\label{sec:introduction}
The interaction of AGN jets with their environments has been under
investigation for several decades, partly because it is easily
observable through the morphology of radio lobes
\citep[e.g.][]{miley:80} and X-ray cavities
\citep[e.g.][]{mcnamara:07}.  Because the most powerful AGNs are
essentially stationary in the centers of galaxy clusters, models of
this interaction typically only consider stationary atmospheres that
jets run into (see, e.g., \citealt{heinz:06b} for recent work on jets
in dynamic atmospheres).  The evolution in this case can be separated
into three distinct phases: (1) the early momentum driven phase, where
the ram pressure of the jet is significant for the dynamical evolution
and the source evolves into an elongated structure with narrow
cocoons, (2) the energy driven phase, where the slowed-down jet plasma
inflates supersonically expanding lobes, excavating a quasi-spherical
cavity (this phase is well described by the self-similar solution by
\citealt{castor:75,kaiser:97,heinz:98}), and (3) the late, sub-sonic
evolution, when the radio lobes (the reservoirs of relativistic gas
released by jets) are in pressure equilibrium with the environment.
In the case of AGNs, this radio plasma is buoyant in the host
galaxy/galaxy cluster atmosphere \citep[see][for a more detailed
  review]{reynolds:02}.

It is becoming increasingly clear that X-ray binaries (XRBs) are also
producing relativistic jets over a wide range in accretion rate
\citep{fender:00,fender:01,gallo:03,fender:04}.  The inner regions of
these XRB jets, where their dynamics is governed by the atmosphere of
the compact object powering them, are very similar to the inner
regions of AGN jets \citep{heinz:03a}.  However, \cite{heinz:02c}
showed that some critical differences exist between XRB jets and AGN
jets in how they interact with the larger scale environment, well
outside of the sphere of influence of the central compact object.  One
of the differences is that, in comparison with AGN jets, the ISM poses
a much weaker barrier to microquasar jets. This is because the XRB jet
thrust (in other words, the ram pressure delivered by the jet) is much
larger in comparison to the inertial density of the ISM than it is in
the case of AGN jets.

The second fundamental difference between the two cases is, of course,
that XRBs do not reside in the centers of dark matter halos with
stratified gaseous atmospheres.  Instead, they travel through regular
Galactic ISM with some space velocity $v_{\rm XRB}$, set by supernova
kicks and orbital dynamics.  The velocity dispersion of XRBs implies
that a significant fraction of these sources are moving with large
(supersonic) velocities through the ISM, which will have important
consequences for their dynamics.  VLBI parallax measurements have
shown that a microquasars can be moving with velocities in excess of
$100\,{\rm km\,s^{-1}}$ with respect to the local standard of rest
\citep{mirabel:01}.  As a result, the ram pressure of the ISM will act
on the radio plasma released by the source, sweeping back the outer
layers of the radio lobe.

In this paper, we argue that this aspect has important consequences
for the interaction of some XRB jets with their environment: Instead
of inflating stationary cocoons, in many cases they will produce
trails of radio emitting plasma (made up of relativistic particles and
magnetic fields) as they travel through the Galaxy.  A very similar
situation is encountered in pulsar bow shock nebulae
\citep{cordes:96,frail:96}, where a relativistic wind inflates a
channel in the ISM through which the pulsar is moving, and, of course,
in exragalactic head-tail sources, which are the direct equivalent of
jets propagating into a moving medium.

The paper is organized as follows: In \S\ref{sec:dynamics}, we will
present a simple parametric model of the interaction of the jets with
the ISM.  In \S\ref{sec:discussion} we discuss the consequences for
particle and magnetic field input into the ISM,
\S\ref{sec:observations} discussed the observational signatures of
this interaction, and \S\ref{sec:summary} presents a brief summary of
the paper.

\section{The interaction of microquasar jets with their environment}
\label{sec:dynamics}
No detailed, global models of the dynamical interaction of XRB jets
with the ISM exist as of yet.  In the absence of such models, the
first logical step is to explore a relatively simple sketch of how
this interaction occurs.  The predictions we will make are based on
what we know about the interaction of AGN jets with their environment,
some insight gained from the scaling relations derived in
\cite{heinz:02c} and on the estimates of the kinetic energy output
from these source derived in \citet{fender:05b} and \citet[][HG05
  hereafter]{heinz:05c}.

Throughout the paper, we will assume that the dynamical time of the
large scale evolution of the source is long compared to the
variability time scale of the XRB and the jet (see \S\ref{sec:early}).
This will allow us to neglect the variable nature of the jet power in
the long term evolution of the large scale radio lobe and instead use
a time averaged constant value.

Based on the estimated power $W_{\rm cyg}\sim 10^{37}\,{\rm
  ergs\,s^{-1}}$ of the jet in Cyg X-1 from \cite{gallo:05}, and
loosely guided by the limit on the average jet power of $\langle
W\rangle \sim few\times 10^{37}\,{\rm ergs\ s^{-1}}$ from
HG05\footnote{The average power in HG05 was derived from a limited
  luminosity range, excluding sources below the luminosity
  completeness limit, inclusion of which would bring the average
  down.}, we adopt a fiducial ensemble average kinetic jet power {\em
  per source} of
\begin{equation}
  \scrW \equiv \langle W\rangle \equiv 10^{37}\,{\rm
    ergs\,s^{-1}}\,\cdot\,W_{37}
\end{equation}

We will also use the estimated average aggregate power from {\em all}
XRB jets of $W_{\rm tot} \sim 5 \times 10^{38}\,{\rm ergs\,s^{-1}}$
This estimate does {\em not} include the even more uncertain
contributions from neutron star jets or optically thin radio
outbursts, both of which will increase the total power and thus the
importance of the effects discussed here.

We will further assume that the gas is sufficiently ionized or
sufficiently dense to treat the magnetized plasma released by the jet
and its environment as fluids and use hydrodynamic arguments (see
\S\ref{sec:mixing}).

We will use a set of fiducial parameters for the ISM.  Since the ISM
itself is turbulent and inhomogeneous we cannot simply assign a mean
density.  We will discuss the three putative main phases (cold, warm,
and hot) separately whenever appropriate.  The ISM pressure is likely
dominated by turbulent pressure and contains contributions from
thermal and non-thermal particles and magnetic fields.  We will use a
fiducial mean total ISM pressure of \citep{cox:05}
\begin{equation}
  p_{\rm ISM} \approx 3 \times 10^{-12}\,{\rm ergs\,cm^{-3}}\,\cdot\,p_{-11.5}
\end{equation}

We will assume throughout the paper that the working surface, i.e.,
the place where jet and ISM interact directly, slowing down the jet
material (taken to be a strong terminal shock\footnote{There is little
  fundamental difference if instead the jet material is decelerated
  through a series of weak shocks along the jet}) is efficient at
slowing the jets down to sub-relativistic velocities and thus
dissipating most of the kinetic energy into internal energy of the
radio plasma.  For a jet with length $l_{\rm jet}$ and half opening
angle $\alpha \equiv 1^{\circ}\alpha_{1}$ (where $\alpha_1$ is the
opening angle measured in degrees), this is the case for
\begin{equation}
  l_{\rm jet} \gtrsim 10^{16}\,{\rm
  cm}\,\cdot\,\sqrt{\frac{W_{37}}{n_{1}}}\frac{1}{\alpha_1}
\end{equation}
where we define the ISM particle density as $n_{1} \equiv n_{\rm
  ISM}/1\,{\rm cm^{-3}}$.  The size scales we are interested in are
much larger than this, so we are justified in assuming the working
surface dissipates most of the kinetic energy.  We will also assume
that this plasma does not mix with the ISM due to magnetic flux
freezing (see, however, \S\ref{sec:mixing}).

\begin{figure}
\begin{center}
\resizebox{0.7\columnwidth}{!}{\includegraphics{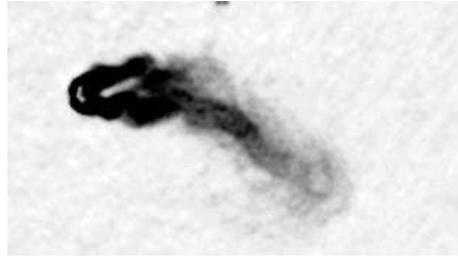}}
\end{center}
\caption{Radio map of extragalactic head-tail source 3C83.1 (NGC
  1265), rotated by 90 degrees \citep{odea:86}.
  \label{fig:3c83}}
\end{figure}

\begin{figure}
\begin{center}
\resizebox{0.7\columnwidth}{!}{\includegraphics{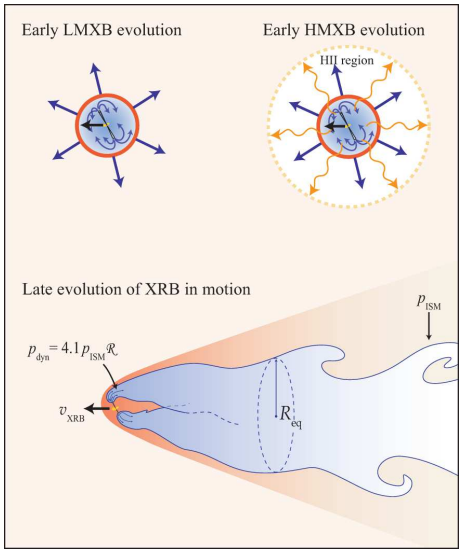}}
\end{center}
\caption{Sketch of early and late evolutionary phases in XRB radio
lobe dynamics. Note that in the early stages the expansion velocity of
the plasma is much faster than the space velocity of the XRB, thus the
evolution is quasi-spherical. Later, the ram pressure due to the XRB
space velocity and the thermal pressure of the ISM dominate, and the
stripped plasma leaves a trail behind the XRB.
  \label{fig:sketch}}
\end{figure}

\subsection{Early evolution}
\label{sec:early}
During the early lobe evolution the ISM pressure $p_{\rm ISM}$,
buoyancy forces, and the ISM ram pressure $\rho_{\rm ISM}v_{\rm
  XRB}^2$ due to the space velocity $v_{\rm XRB}$ of the XRB are
negligible compared to the ram pressure $\rho_{\rm ISM}v_{\rm R}^2$ on
the lobe due to its radial expansion velocity $v_{\rm R}$. It is then
appropriate to use the well-known 1-dimensional energy-driven bubble
model \citep{castor:75,kaiser:97,heinz:98,gallo:05} as sketched in
Fig.~\ref{fig:sketch}.  This model assumes that the XRB releases
relativistic plasma in a spherically symmetric fashion\footnote{This
  assumption is motivated by the fact that the internal sound speed in
  a relativistic plasma is much larger than the expansion velocity of
  the radio lobe.  Thus, regions of overpressure (i.e., the working
  surfaces of the jets, where the jets interact with the ISM) will
  quickly distribute their overpressure within the relativistic plasma
  cocoon \citep[e.g.][]{begelman:89}.  Furthermore dynamical
  instabilities and precession are believed to spread the jet thrust
  out over a much larger solid angle than subtended by the opening
  angle of the jet for the case of AGN jets \citep{scheuer:82},
  driving the aspect ratio of lobes toward unity. We will assume that
  the effective half opening angle is $\theta=\sqrt{\Omega/\pi} \sim
  20^{\circ}$ \citep[e.g.][]{young:05} where $\Omega$ is the solid
  angle swept out by precession and jitter of the jet axis.}

Assuming (a) an adiabatic equation of state, (b) uniform pressure for
the radio plasma inside the lobe, (c) ram pressure balance with the
ISM, and (d) energy conservation, the Castor solution is
\begin{equation}
  R_{\rm early} \approx 0.65 \left(\frac{\langle W \rangle
      t^3}{\rho_{\rm ISM}}\right)^{1/5}
\label{eq:early}
\end{equation}

Apart from the expected synchrotron emission from inside the radio
lobe, the supersonic expansion into the ISM will also lead to thermal
emission from the shocked ISM shell \citep{kaiser:04}, similar to
expanding supernova remnants.  Shocks are extremely valuable as
diagnostic tools, since the shock strength measures the expansion
velocity of the source, which, when coupled with the source size, can
tell us the age and average kinetic power of the source.  An example
of such a source is Cyg X-1, for which this thermal emission has been
detected \citep{gallo:05}.

\subsection{Late evolution: Head-tail sources}

At sufficiently late times (the case we are interested in here) the
jets will have traveled sufficiently far for the dynamical pressure
due to the XRB motion to become important.

The dynamical pressure of the XRB motion through the ISM with velocity
$v_{\rm XRB}$ will become comparable to the jet thrust (i.e., the
momentum flux along the jet) when the advance speed of the working
surface equals the space velocity of the source, $v_{\rm head}=v_{\rm
  XRB}$.  If this condition is satisfied, the working surface can no
longer advance in the direction of the jet and will thus become
stationary in the frame of the XRB.  Given our fiducial sound speed of
$c_{\rm sound}=17\,{\rm km\,s^{-1}}$ we will express the XRB's space
velocity as
\begin{equation}
  v_{17}\equiv \frac{v_{\rm XRB}}{17\,{\rm km\,s^{-1}}}
\end{equation}

The shock jump conditions and pressure balance across the working
surface (the terminal shock) imply that this happens when the jet
thrust
\begin{equation}
  p_{\rm jet} = \frac{5\Gamma +3}{\Gamma v_{\rm
  jet}}\frac{W_{\rm jet}}{12 \pi(\theta l)^2}
\end{equation} 
equals the total dynamic pressure
\begin{eqnarray}
  p_{\rm dyn} & = & 4.1\,p_{\rm
    ISM}\,\mathcal{R}_{M} \label{eq:pdyn}\\ \mathcal{R}_M & \equiv &
  \left\{9 + 10M^2 \left[1 + \sqrt{1+9/(4M^2)}\right]\right\}/37
  \nonumber
\end{eqnarray}
of the ISM, where $M \equiv v_{\rm XRB}/c_{\rm sound}$ is the
effective Mach number of the XRB's motion through the ISM with respect
to the effective sound speed of the ISM, $c_{\rm sound}=\sqrt{5p_{\rm
    ISM}/3\rho_{\rm ISM}}$.  We adopt a fiducial value of $\Gamma
\equiv 5\Gamma_{5}$ for the bulk Lorentz factor of the jet, which is,
however, of little importance throughout the rest of the paper.

The expression for $p_{\rm dyn}$ in eq.~(\ref{eq:pdyn}) can easily be
derived from the shock jump conditions on each side of the working
surface \citep[e.g.][]{blandford:76}, with non-relativistic equations
of state for the unshocked cold jet and ISM gas\footnote{The
  $\gamma=5/3$ equation of state is not a crucial assumption and can
  easily be modified to reflect a relativistic equation of state of
  the jet gas (for example).  The coefficients only change slightly
  and the qualitative results are unaltered.}.

The two dynamical pressures $p_{\rm dyn}$ and $p_{\rm jet}$ are equal
when the length of the jets reaches
\begin{eqnarray}
  l_{\rm jet} & = & \frac{6 \times 10^{18}\,{\rm
      cm}}{\theta_{20}}\,\cdot\,\left[\frac{W_{37}}{p_{-11.5}}\frac{25\Gamma
      + 15}{28\Gamma\beta\mathcal{R}_M}\right]^{1/2}
  \label{eq:length}
\end{eqnarray}
where the expression in brackets is unity for $M=1$ and the fiducial
parameters and $\beta=v_{\rm jet}/c$.  At this length, the working
surface becomes stationary in the frame of the XRB, and since the ISM
is moving at velocity $v_{\rm XRB}$ with respect to the XRB, the
plasma that is released at the working surface must be swept back away
from the XRB with the ISM at velocity $v_{\rm XRB}$ and create a trail
of ``debris'' plasma, left behind by the XRB jets.

The time for the jet to propagate that far into the ISM is
approximately $t_{\rm eq} \approx l_{\rm jet}/v_{\rm XRB} \approx
10^{5}\,{\rm yrs}$, which is long compared to typical accretion disk
time scales \citep[e.g.][]{frank:00} but still short compared to
typical companion (i.e., XRB) lifetimes.  It is therefore necessary to
further discuss this quasi stationary phase of the large scale
dynamics of radio plasma released by XRBs (it is quasi stationary
since the XRB space velocity $v_{\rm XRB}$ and the local ISM density
$n_{\rm ISM}$ change as the XRB traverses the Galaxy).

\subsection{The debris: Radio trails}

As mentioned above, we assume that the working surface is stationary
in the frame of the XRB.  Its propagation velocity into the ISM must
therefore be $v_{\rm XRB}$ and its pressure must be given by $p_{\rm
  dyn}$ from above.

The energy released by the jet at the working surface must then equal
the jet power:
\begin{equation}
  W_{\rm jet}=\frac{\gamma}{\gamma - 1}p_{\rm dyn}\frac{dV_{\rm
      head}}{dt}
\end{equation}
where $V_{\rm head}$ is the volume of plasma streaming out of the
working surface.  The factor $[{\gamma_{\rm ad}}/({\gamma_{\rm
    ad}-1})]p_{\rm dyn}$ is the post-shock enthalpy of the jet plasma
and includes both the internal energy $pdV/({\gamma_{\rm ad}-1})$ of
the plasma released and the $pdV$ work done on the ISM.  Since
$\gamma_{\rm ad}=4/3$ for the jet plasma, the volumetric rate at which
jet plasma is released at the working surface is
\begin{equation}
  \frac{dV_{\rm head}}{dt} = \frac{W_{\rm jet}}{4\,p_{\rm dyn}}
\end{equation}

Since far downstream (along the swept back trail of radio plasma away
from the source) the pressure has to equal the ISM pressure (otherwise
the energy in the Mach cone behind the XRB would diverge), the radio
plasma originally released at pressure $p_{\rm dyn}$ must expand
adiabatically (thus doing further work on the ISM).  Thus, the
asymptotic volumetric rate at which plasma enters the trail is
\begin{equation}
  \frac{dV}{dt} \approx \frac{W_{\rm jet}}{4\,p_{\rm
      dyn}}\left(\frac{p_{\rm dyn}}{p_{\rm
      ISM}}\right)^{\frac{1}{\gamma_{\rm ad}}} = \frac{W_{\rm
      jet}}{4\,p_{\rm dyn}}\left(\frac{p_{\rm dyn}}{p_{\rm
      ISM}}\right)^{3/4}
  \label{eq:dvdt}
\end{equation}

In order to estimate the asymptotic lateral radius $R_{\rm eq}$ of the
trail (approximated as a cylinder around the XRBs trajectory) from its
inflation rate (eq.~\ref{eq:dvdt}), we need to know the asymptotic
flow velocity $v_{\rm tr}$ of radio plasma through the trail.
The pressure gradient behind the bow shock (from the ISM ram pressure
$p_{\rm dyn}$ to $p_{\rm ISM}$) will accelerate the plasma along the
channel.  

In principle, it is possible that the plasma will pick up most of its
original outflow velocity and reach mildly relativistic speeds.  For
example, \cite{bucciantini:05} investigated the dynamics of pulsar bow
shock nebulae using axi-symmetric 2.5D relativistic MHD simulations
and found that the velocity inside the channel reaches values up to
$0.6c$.  They concluded, on the basis of their simulations, that the
channel width should be comparable to the standoff distance between
the pulsar (or XRB in our case) and the stagnation point at the head
of the shock.  However, the enforced symmetry in their model makes the
solution they found over-stable.  In reality, it is likely that
dynamical instabilities (e.g., Kelvin-Helmholtz or kink instabilities,
depending on the strength and topology of the magnetic field) would
disrupt such rapid flow through the channel.

We expect the average flow velocity to be roughly that of the ISM
sheet around the channel.  Asymptotically, this velocity must be
$v_{\rm XRB}$ (otherwise the total kinetic energy would diverge).
Thus, we will assume that the plasma, far downstream from the XRB will
have a flow velocity of $v_{\rm tr} \equiv v_{\rm XRB}\scrv$ with
$\scrv \gtrsim 1$.

Note that even in the case of $\scrv \gg 1$, the plasma must still
come to rest in the ISM at some point, where it would inflate a
bubble.  This would be around the birthplace of the XRB or around any
major change in environmental parameters (i.e., changes in ISM density
or pressure), where the channel width would change significantly,
leading to a series of ``sausage link'' bubbles, connected by narrow
channels.  Note also that the total volume of plasma released should
not be affected.

The equilibrium radius $R_{\rm eq}$ of the roughly cylindrical trail
is then
\begin{eqnarray}
  R_{\rm eq} & \approx & \sqrt{\frac{dV}{dt}\frac{1}{\pi\,v_{\rm
	XRB}\scrv}} \nonumber \\
  & \approx & 2 \times 10^{20}{\rm
	cm}\,\cdot\,\sqrt{\frac{W_{37}} {p_{-11.5}\,v_{17}\,\scrv}}
	\mathcal{R}_M^{-\frac{1}{8}}
  \label{eq:req}
\end{eqnarray}

This is much larger than the value of $l_{\rm jet}$
(eq.~\ref{eq:length}).  Thus, the ultimate cross section of the trails
is much larger than the equilibrium length of the jets and our
assumption that both jets are feeding a single trail is justified
(this is not qualitatively important for our argument, however).

It is worth pointing out the implicit weak dependence of $R_{\rm eq}$
on density through $\mathcal{R}_{M}^{-1/8}$ which implies that the
size scale of the trails produced should be similar in different
environments - an increase in density by three orders of magnitude
(with otherwise identical parameters, i.e., pressure and jet power)
only reduces $R_{\rm eq}$ by a factor of 3.5, while a decrease in
density by three orders of magnitude (i.e., a hot ISM phase
environment) only increases $R_{\rm eq}$ by a factor of 1.3.  Thus,
the most significant effects on $R_{\rm eq}$ are likely to come from
$W_{37.3}$, $p_{-11.5}$, and $v_{17}$.

For a binary lifetime of $t \equiv t_{8} \times 10^8\,{\rm yrs}$
(where $t_{8}$ is the age of the source in units of $10^8$ yrs), the
length of the trail left by the source is
\begin{equation}
  l_{\rm trail}=5\times 10^{21}\,{\rm cm}\,\cdot\,v_{17}\,t_{8}
\end{equation}
Based on this expression, a source will produce a significant plasma
trail if $l \gg R_{\rm eq}$, or
\begin{equation}
  t_8 p_{-11.5}^{1/2} v_{17}^{3/2}\scrv^{1/2} W_{37}^{-1/2}
  \mathcal{R}_M^{\frac{1}{8}} \gg 0.06
  \label{eq:trailcondition}
\end{equation}
If this condition is not satisfied, the source dynamics are better
approximated as spherical and stationary (if significant gradients
exist in the environment of the XRB, the source evolution will, of
course, still deviate from a quasi-spherical, symmetric case, as might
be the case in Cygnus X-1).

The fiducial values used in eq.~(\ref{eq:trailcondition}) are
appropriate for LMXBs, implying that LMXBs typically leave trails in
the ISM, especially the high velocity tail of the LMXB population with
$v_{17} \gg 1$.  Since HMXBs have low space velocities and short
lifetimes, this condition is typically {\em not} satisfied for HMXBs:
For a characteristic HMXB velocity of $v_{17} \scrv \sim 0.35$ and a
companion lifetime of $t_8 \lesssim 0.1$, the expression on the left
is $0.015$.  Thus, HMXBs typically should {\em not} produce trails,
except for very low power source, and sources in the very high
velocity tail of the HMXB distribution in dense, high pressure
environments (such as the Galactic center).

In arguments above we have implicitly assumed that a fluid treatment
is valid, i.e., that the relativistic particles released by the jet
interact collectively with the ISM.  This is equivalent to assuming
that the particles are frozen to the plasma and that microscopic
mixing between ISM and trail plasma is not a dominant effect.  We will
discuss this assumption and its validity in more detail in
\ref{sec:mixing}.  Any significant amount of mixing will only make the
conclusions we will draw below about seeding of magnetic fields into
the ISM stronger (however, it might alter the observation appearance
of trails as well).

\section{Mixing and the impact on the ISM}
\label{sec:discussion}
Given the estimates for the properties of the radio lobes and trails
produced by the ensemble of Galactic XRBs, we can estimate their
impact on the ISM.  Before going into detail, we will lay out our
assumptions about the ISM that the XRBs are interacting with:

For the XRB velocities, we will use a Maxwellian distribution with
mean velocities of $\sigma_{\rm H} \approx 6\,{\rm km\,s^{-1}}$ and
$\sigma_{\rm L} \approx 16\,{\rm km\,s^{-1}}$ for HMXBs and LMXBs
respectively, which reproduces the scale heights of both populations.
A fraction of about $f_{\rm disk} \sim 75\%f_{\rm disk,75}$ of the
LMXBs reside in the Galactic disk.

For the ISM density distributions we will use volume filling fractions
of $f_{\rm hot} \approx 70\%$, $f_{\rm warm}\approx 25\%$, and $f_{\rm
  cold} \approx 5\%$ for the hot, warm, and cold phases respectively
\citep{mckee:77}.  We will assume that LMXBs are randomly distributed
among the three phases since they are old enough to have moved far
enough away from their birth places to have forgotten about their
initial environments.  HMXBs do not live long enough and have
velocities too small to have traveled very far from their birth place,
about 60\,pc for $v_{\rm XRB}=6\,{\rm km/s}$ and a life time of
$10^{7}\,{\rm yrs}$.  This has to be compared to typical molecular
cloud size of about 100\,pc.  We will thus assume that $f_{\rm
  cold}=100\%$ for HMXBs to be conservative (smaller values of $f_{\rm
  cold}$ will lead to larger values of the total volume of plasma
released by HMXBs below).

\subsection{The total volume of radio plasma released into the
  Galaxy}

The total volume in radio trails produced by XRBs depends on the
distribution of $v_{\rm XRB}$ and $\rho_{\rm ISM}$ only through
$\scrR^{1/4}$.  Thus, estimates of the volume of plasma released by
jets will be robust against uncertainties in those quantities.

With the ISM parameters given above, the volume in trails produced by
the LMXB disk population is simply the current rate of volume
injection per LMXB, averaged over the velocity distribution of LMXBs,
multiplied by the number of LMXBs, multiplied by the age of the LMXB
population $t_{\rm L} \equiv \times 10^{10}t_{10}$ yrs. 

Using the estimated total integrated kinetic power of the LMXB
population of $W_{\rm L} \approx 3.5\times 10^{38} W_{\rm L,38.5}$,
this gives:
\begin{eqnarray}
  V_{\rm LMXB,disk} & \approx & W_{\rm L} t_{\rm L} \langle
    \left(p_{\rm dyn}/p_{\rm ISM}\right)^{3/4}/ (4 p_{\rm dyn})
    \rangle \nonumber \\ & \approx & 5.5 \times 10^{66}\,{\rm
    cm^{3}}\,\cdot\,\frac{W_{\rm L,38.5}f_{\rm
      disk,75}t_{10}}{p_{-11.5}}\nonumber \\
    & & \ \ \ \ \ \cdot h_{\rm L}(\sigma_{\rm L},f_{\rm
    cold,warm,hot}) \label{eq:vollmxb}
\end{eqnarray}
where $h_{\rm L}$ is unity for our fiducial parameters and a very
slowly varying function of $\sigma_{\rm L},\ n_{1}$, and $\ f_{\rm
  cold,warm,hot}$ over the parameter range considered.  The velocity
dispersion $\sigma_{\rm L}$ drops out of eq.~(\ref{eq:vollmxb}) to
lowest order.  Clearly, the most uncertain parameter left is $W_{\rm
  LMXB,38.5}$, on which $V_{\rm tot}$ depends linearly.

Taking $f_{hot}=100\%$ for the halo population, the volume produced by
the halo LMXBs is about $V_{\rm L,halo} \approx 2 \times 10^{66}\,{\rm
  cm^{3}}/p_{-11.5}$.  However, a sizeable fraction of the plasma
produced in the disk will eventually leak into the halo (see below).
Since $p_{-11.5}$ is much smaller in the halo than it is in the disk,
the total volume filled by plasma released by halo LMXBs and that
emanating buoyantly from the disk (see below) is thus likely much
larger than this estimate.

For the HMXBs, the total volume rate should be integrated over the
star formation rate history, assuming a current star formation rate of
$3\,M_{\odot}\,{\rm yr^{-1}}\,\dot{M}_{3}$ (where $\dot{M}_{3}$ is the
Galactic star formation rate in units of 3 solar masses per year).

With the slightly lower value of the total power $W_{\rm H}$ and the
total star formation rate integral of $\int dt \dot{M}_{\rm SFR} =
M_{\rm Gal} \equiv 10^{11}\,M_{\odot}\,M_{11}$ (where $M_{11}$ is the
Galactic stellar mass in units of $10^{11}$ solar masses), this gives
\begin{eqnarray}
  V_{\rm H,disk} & \approx & 4\times 10^{66}\,{\rm
    cm^{3}}\,\cdot\,\frac{W_{\rm H,38.3} M_{11}}{p_{-11.5}\dot{M}_{3}}
  h_{\rm H}(\sigma_{\rm H},f_{\rm c}) \label{eq:volhmxb}
\end{eqnarray}
Again $h_{\rm H}$ is slowly varying.

Summarizing, the total volume of radio plasma released into the disk
by XRB jets over the lifetime of the Galaxy is then roughly $8\times
10^{66}\,{\rm cm^{3}}$, and the volume released by halo LMXBs is
roughly $2 \times 10^{66}\,{\rm cm^2}\times (p_{\rm disk}/p_{\rm
  halo})$.

This should be compared to the disk and the halo volumes: Taking the
disk scale height to be 500 pc and the disk radius to be 10 kpc, the
disk volume is $V_{\rm disk} \approx 10^{67}\,{\rm cm^{3}}$, while the
halo volume is roughly $V_{\rm halo} \gtrsim 10^{68}\,{\rm cm^{3}}$.
Clearly, XRBs will have released sufficient relativistic plasma to
fill a significant fraction of the disk and even the halo. If the
radio plasma were confined to the disk, it would fill up to 100\% of
the disk (and could account for all of the hot ISM phase in the disk).
This immediately implies that the buoyancy of this plasma in the
vertically stratified Galactic gas will eventually move most of the
trails out into the halo.  It also implies that a volume fraction
larger than 10\% of the halo might be filled with radio plasma
produced by XRBs.  We will discuss further implications of this result
below.

\subsection{Microscopic mixing}
\label{sec:mixing}
Microscopic mixing of radio plasma and ISM is only going to be
important in sufficiently neutral ISM phases, since otherwise the
strong magnetization of the trail plasma will effectively decouple
both phases.

While the ISM will be sufficiently ionized near the working surface,
it might well be neutral far back ablong the trail.  In this case,
neutral particles can enter the trail. The cross section for
collisional interaction between the neutral particles and the cosmic
rays of the trail is negligible.  Thus, the only way for neutral ISM
particles to interact at all with the trail plasma is through
interaction with ionized ISM particles via ambipolar diffusion and
charge exchange reactions (we will ignore the latter given the
expected low cross sections).

While a detailed discussion of microscopic mixing of relativistic
plasma across radio lobe boundaries is beyond the scope of this paper,
we will briefly present an order of magnitude estimate of the
importance of ambipolar diffusion of neutral ISM into XRB radio lobes
and trails to show that the exptected amount of mixing is
astrophysically interesting.

Following \cite{draine:83}, we use a constant ion drag coefficient of
$\gamma_{\rm ion}=3.5\times 10^{13}\,{\rm cm^{3}\,g^{-1}\,s^{-1}}$.
In the cold neutral phase, it is appropriate to use an ionization
fraction of $\xi_{\rm i}=n_{\rm i}/n_{\rm n}=C_{\rm i}\rho_{\rm
  n}^{-1/2}$ with $C_{\rm i} \approx 3\times 10^{-16}\,{\rm
  g^{1/2}\,cm^{-3/2}}$ \citep{elmegreen:79}.  In the warm neutral
medium, the ionization fraction is higher, of order $\xi_{\rm
  i}\approx 10\%$ \citep[e.g.][]{dalgarno:72}.

Next, we need to estimate the field strength inside the trail.  Given
the estimated pressure and size of the plasma trail, we can
parameterize the field strength in terms of its equipartition value,
\begin{eqnarray}
  B_{\rm eq} & = & \sqrt{3 \xi_{B} 8 \pi\,p_{\rm
      trail}/(1+\xi_{B})} \label{eq:bfield} \\ & \approx & 11\,{\mu\rm
    G}\,\cdot\,\sqrt{2\xi_{\rm B}p_{-11.5}/(1 + \xi_{\rm B})}\nonumber
\end{eqnarray}
where $\xi_{B}$ is the ratio of magnetic to particle pressure and
$p_{\rm tot}$ is the total (particle plus magnetic) pressure.  For
simplicity, we assume the field is tangled isotropically, leading to
the additional factor of $3$ \citep[e.g.][]{heinz:00}.

The ambipolar diffusion coefficient for cold neutrals entering the
trail/lobe is then \citep[e.g.][]{shu:92}
\begin{eqnarray}
  \lefteqn{D_{\rm amb,cold}  =  \frac{v_{\rm A}^2}{\gamma\rho_{\rm
      i}}=\frac{B^{2}}{4\pi\rho_{\rm n}}\frac{1}{\gamma\,C_{\rm
      i}\,\rho_{\rm n}}} \nonumber \\ & & \approx 4\times 10^{20}\,{\rm
    cm^{2}\,s^{-1}}\,\cdot\,p_{-11.5}\left(\frac{10^{4}\,{\rm cm^{-3}}}{n_{\rm
      n}}\right)^{3/2}\frac{2\xi_{\rm B}}{1 + \xi_{\rm B}}\nonumber
\end{eqnarray}
and for warm neutrals
\begin{equation}
  D_{\rm amb,warm} = 10^{24}\,{\rm
    cm^{2}\,s^{-1}}\,\cdot\,\frac{p_{-11.5}}{n_{\rm
      n}^2}\frac{10\%}{\xi}\frac{2\xi_{\rm B}}{1 + \xi_{\rm B}}
\end{equation}
We are interested in the asymtotic mixing fraction far back along the
trail, when it has reached its equilibrium radius.  Thus, we will not
consider dynamical effects due to the trails initial expansion on the
total mixing fraction\footnote{During the radial expansion phase of
  the lobe and in the radially expanding section of the trail,
  ambipolar diffusion will only work as long as the diffusion speed is
  faster than the expansion velocity}.

A conservative assumption about the longevity of the radio trail is
that it will last at least one sound crossing time against any
dynamical instabilities (see \S\ref{sec:destruction}).  Within a sound
crossing time $\tau_{\rm sound}=R_{\rm eq}/c_{\rm sound}$, the
ambipolar diffusion length is
\begin{equation}
  R_{\rm amb} \approx \sqrt{\frac{D_{\rm amb}}{\tau_{\rm sound}}}
\end{equation}
Thus, a lower limit to the mixing fraction of thermal material from
the cold neutral phase into radio trails/lobes produced by
microquasars is
\begin{eqnarray}
  f_{\rm mix,cold} & \equiv & \frac{V_{\rm mix}}{V_{\rm tot}} \approx
  2\frac{R_{\rm amb}}{R_{\rm eq}} \geq 2\frac{\sqrt{D_{\rm
        amb}\tau_{\rm sonic}}}{R_{\rm eq}}\\ & \geq &
  2\%\,\cdot\,\sqrt{\frac{n_{1}}{10^{4}}}\sqrt{\frac{1 + \xi_{\rm
        B}}{2\xi_{\rm B}}} \left(\frac{W_{37}}
  {p_{-11.5}^{2}v_{17}\scrv}\right)^{1/4} \scrR^{-1/16}\nonumber
\end{eqnarray}
and for warm neutrals
\begin{eqnarray}
  f_{\rm mix,warm} & \geq & 10\%\sqrt{\frac{\xi_{\rm i}}{10\%}}
  \,\cdot\,\left(\frac{n_{1}^{3}W_{37}}
  {p_{-11.5}^{2}v_{17}\scrv}\right)^{1/4} \nonumber \\ & & \ \ \ \ \
  \cdot \scrR^{-1/16}\sqrt{\frac{1 + \xi_{\rm B}}{2\xi_{\rm
  B}}}\nonumber
\end{eqnarray}
We conclude that, for the warm and cold neutral ISM, the magnetic
field inside the trail presents only a moderate barrier and mixing
could be relatively efficient, quickly creating a boundary layer of a
few percent thickness around the trail/lobe.  At the same time,
ambipolar diffusion is probably not strong enough to affect the
dynamics of the radio plasma on a global scale.

In conclusion, mixing is significant in the neutral ISM phases, with
mixing fractions of order of a few percent. It is likely orders of
magnitude smaller for XRBs in ionized environments\footnote{Anomalous
  diffusion can lead to significant mixing even between two completely
  ionized phases \citep[e.g.][]{narayan:01}.}.  This suggest that the
global average of $f_{\rm mix}$ should be about a factor of 10 smaller
than the volume filling fraction of neutral gas in the Galaxy.  So,
given recent estimates of volume filling factor of up to 40\% for the
warm neutral phase \citep{heiles:03} we ``guesstimate'' a globally
averaged mixing fraction of order $f_{\rm mix} \lesssim 4\%$.

\subsection{Competing destruction processes}
\label{sec:destruction}
As the radio plasma expands into the ISM a number of processes will
compete to either mix the two phases or remove the plasma from the
Galactic disk into the halo.

\subsubsection{Buoyancy} 
The ISM is vertically stratified due to the gravitational field of the
Galaxy. The low density radio trails are buoyantly unstable in this
gas, which will drive them upward, away from the Galactic plane.

The buoyancy time scale for a cylindrical trail of radius $R$ is set
by the downward ISM dynamical pressure and the upward buoyancy of the
radio plasma.  For a Galactic rotation velocity $v_{\rm r} \sim
220\,{\rm km\,s^{-1}}$ (which sets the vertical gravity $g_{\rm z}$),
a Galacto-centric distance of $r \sim 10\,{\rm kpc}\,r_{10}$ (where
$r_{10}$ is the Galacto-centric distance in units of $10\,{\rm kpc}$),
and a vertical trail position $z \sim 200\,{\rm pc}\,z_{0.2}$ above
the Galactic mid-plane (where $z_{0.2}$ is the trail height in units
of 200\,pc), and assuming a drag coefficient $C_{\rm W}$ of about 1
for a cylindrical trail, the buoyancy speed is
\begin{eqnarray}
   v_{\rm B} & \approx & \sqrt{\frac{g_{\rm z}\pi R}{C_{W}}} \approx
   \frac{v_{\rm rot}}{r}\sqrt{{\pi R z}} \\ & \approx & 3 \times
   10^{5}\,{\rm cm\,s^{-1}}\,\cdot\,\frac{z_{0.2}^{1/2}}{r_{10}}
   \left(\frac{W_{37}}{p_{-11.5}v_{17}}\right)^{1/4}{\scrR^{-1/16}}
   \nonumber
\end{eqnarray}

For a disk scale height of $H\approx 500\,{\rm pc}\, H_{0.5}$, the
buoyancy time is
\begin{eqnarray}
   \tau_{\rm b} & \sim & H/v_{\rm B} \\ 
   & \approx & 1.6\times 10^{8}\,{\rm
   yrs}\,\cdot\,\frac{r_{10}\,H_{0.5}}{z_{0.2}^{1/2}}
   \left(\frac{p_{-11.5}v_{17}}{W_{37.3}}\right)^{1/4}\scrR^{-1/16}
   \nonumber
\end{eqnarray}
which is, be default, longer than a sound crossing time (since
buoyancy is always sub-sonic).  This is comparable to the lifetime of
an HMXB, but could be short compared to the lifetime of an LMXB.

Unhindered, this process will eventually remove all of the radio
plasma from the disk into the halo.  However, a number of processes
will work against this transport out of the plane, trying to mix the
relativistic plasma into the ISM.  In order to estimate their
importance, one needs to compare the time scales over which they act
to the buoyancy time.  If they are occurring on more rapid time
scales, they might mix a significant amount of radio plasma
macroscopically with the ISM before it has time to escape into the
halo, especially because all of these processes will only increase the
trails surface area and thus its buoyancy time.  At the same time, an
increase in surface area will also increase the microscopic mixing
fraction $f_{\rm mix}$.

\subsubsection{Shredding}
Due to the differential rotation of the Galaxy, any object of finite
size will experience Galactic shear.  Since the radio trails are
embedded in the ISM, they will be affected.

Taking the Galactic rotation velocity $v_{\rm r} \sim 220\,{\rm
  km\,s^{-1}}$ to be constant with radius, the angular velocity shear
across the trail is $d\omega/dr = -v_{\rm r}/r^2$.  The Shredding time
for an object much smaller than its Galacto-centric distance $r$ that
is subject to this shear is
\begin{equation}
  \tau_{\rm shear} \sim r/v_{\rm r} \sim 5\times 10^{7}\,{\rm
    yrs}\,\cdot\,r_{10}
\end{equation}
independent of the actual size of the object.

After several shredding times the object will have been significantly
stretched and distorted and will eventually be mixed with the
ISM. Note that the shredding time is approximately equal to the
buoyancy time, so for an XRB in the Galactic plane, it is plausible
that shredding will distort, and possibly even destroy the radio trail
before it can escape into the halo.

\subsubsection{RT and KH instabilities}
In addition to shear-shredding, the trails will also be subject to
dynamical instabilities.  The same buoyancy process that drives the
trails out of the Galactic plane will also trigger the formation of
Rayleigh-Taylor instability, acting to destroy the trail and
macroscopically mix the radio plasma with the ISM.  The growth time
for a mode with wave number $k$ is $\tau \sim \sqrt{gk}$ (for large
density contrast between trail and ISM).  Since the trails are filled
with magnetized plasma, magnetic tension will suppress the growth of
small wavelength modes.  However, global modes will likely not be
suppressed by tangled fields (at least in the linear regime), and so
the growth time for a body mode with wave number $k= 2\pi/R_{\rm eq}$
will be
\begin{equation}
  \tau_{\rm RT} \sim \sqrt{\frac{2\pi g}{R}} \sim \frac{r}{v_{\rm
  r}}\sqrt{\frac{R}{2\pi z}} = \tau_{\rm
  b}\frac{R}{H}\sqrt{\frac{1}{4C_{W}}} < \tau_{\rm b}
  \label{eq:rt}
\end{equation}
with $C_{W} \sim 1$ being the drag coefficient.  Thus, RT instability
will set in well before buoyancy had time to remove the plasma out of
the disk and into the Halo.  If RT instability does grow at the
hydrodynamic rate, it will fragment the radio trails into smaller
pockets, which will be subject to further RT instability.

Eq.~(\ref{eq:rt}) shows that RT instability is more efficient the
smaller the size of the plasma pocket, thus, {\em if} RT instability
is not suppressed on scales comparable to $R_{\rm eq}$ by the presence
of large scale magnetic fields, it will destroy the radio trail into a
spectrum of bubbles small enough for magnetic tension to stabilize
them before buoyancy transports the plasma out of the disk.  Since the
buoyancy time of smaller bubbles is longer, the RT fragmentation will
delay buoyant transport and give other processes more time to act and,
by increasing the surface area of the plasma, increasing the
efficiency of diffusive processes.

In addition to RT instability, Galactic differential rotation and
buoyancy drift will induce KH instability along the edge of the trail.
The typical velocities encountered are of order $v_{\rm buoy} \sim
v_{\rm r}\sqrt{Rz}/r$ and $v_{\rm shear} \sim 2v_{\rm r}R/r$.  $v_{\rm
  shear}$ will dominate at early times, $v_{\rm buoy}$ will become
larger after a few buoyancy times.  Taking the relativistic inertia of
the radio plasma into account, the KH growth time of a body mode with
$\lambda \sim R$ is roughly
\begin{eqnarray}
  \tau_{\rm KH} & \sim & \sqrt{\frac{R^2}{v_{\rm bouy}^2 + v_{\rm
        shear}^2}\frac{\rho_{\rm ISM}c^2}{4p_{\rm ISM}}} \\ & \sim &
  R\tau_{\rm b}\tau_{\rm shear}\sqrt{\frac{5}{12(H^2\tau_{\rm shear}^2 +
      z^2\tau_{\rm b}^2)}}\frac{c}{c_{\rm sound}}
\end{eqnarray}
Clearly, the ratio $c/c_{\rm sound}$ is much larger than either $z/R$
or $H/R$ for typical trail radii (see eq.~\ref{eq:req}), and KH
instability should not contribute much to the global destruction of
radio bubbles and trails.  However, it could well induce a spectrum of
short wavelength modes that creates a shear layer around the trail,
where efficient mixing between the two phases occurs.

\subsection{The impact on the ISM: Magnetization and cosmic ray seeding}
Given the total volume of plasma released by XRBs and some value of
$f_{\rm mix}$, we can estimate the strength of the magnetic field that
is released and seeded into the ISM.

We will use the expression for the field strength inside the trail
from eq.~(\ref{eq:bfield}).  Further assuming magnetic flux
conservation, the strength of the magnetic field after the residual
fraction $f_{\rm mix}$ of the radio plasma is mixed with the ISM is
\begin{eqnarray}
  B_{\rm mixed} & \approx & f_{\rm mix} B_{\rm eq} \\ & \sim &
  11\,f_{\rm mix}\,{\mu\rm G}\,\cdot\,\sqrt{\frac{2\xi_{\rm
        B}}{p_{-11.5}\left(1 + \xi_{\rm B}\right)}} b \nonumber \\ b & \equiv &
  \frac{5.5W_{\rm L,38.5}f_{\rm disk,75}t_{10}h_{\rm L} + 4W_{\rm
      H,38.3}M_{11}h_{\rm H}/\dot{M}_{3}}{9.5} \nonumber
\end{eqnarray}
where $b$ is unity for the set of fiducial parameters.  Since $t_{10}$
and $M_{11}$ enter linearly into this expression, the magnetic field
provided by XRB jets will have been linearly increasing as a function
of time for most of the Galactic history, but will have been
increasing more rapidly during intense star formation episodes, when
$M$ increased dramatically, since then the HMXB population would have
been much more numerous.

This should be compared to an estimated mean Galactic field of order
$B_{\rm G} \sim 10\,{\mu\rm G}$ \citep[e.g.][]{wentzel:63}, which is
believed to have been created by dynamo action on a small seed
magnetic field.  Comparing both values implies that $f_{\rm
  mix}\sqrt{2\xi_{\rm B}/[p_{-11.5}(1 + \xi_{\rm B})]} b < 1$, which
is easily satisfied since we estimated $f_{\rm mix} \lesssim 0.04$
above.

The seed field required to produce the observed Galactic field is
rather small \citep[$B_{\rm seem} \sim 10^{-17}{\rm
    G}$][]{anderson:92}.  Even under the most conservative assumptions
possible --- mixing of only one skin depth of the thickness of one
proton gyro radius ($f_{\rm mix} \sim 10^{-9}$) --- such a field
strength could have been provided by XRB jets within about $10^5\,{\rm
  yrs}$ after the first XRBs turned on (though it would need a few
Galactic revolutions of duration $\tau_{\rm rev} \sim 2.5 \times
10^{8}\,{\rm yrs}$ to get spread over a significant fraction of the
disk).  More reasonable choices of $f_{\rm mix}$ imply that the
strength of the seed field provided by XRB jets is likely orders of
magnitude larger.  In fact, given the numbers derived above, is likely
that the strength of the field seeded by XRB jets is only about two
orders of magnitude below the observed mean Galactic value.

Of particular importance in this case is the fact that LMXBs travel
far from their birthplaces, leaving behind trails of magnetized
plasma, which should permeate the Galactic plane more or less
uniformly (modulated by the average space density as a function of
Galacto-centric distance).

Since XRB production is an unavoidable part of star formation
\citep{grimm:02} and since jet activity is an integral part of XRB
activity \citep[e.g.][]{fender:01c}, it is clear that jets must play a
role in providing magnetic fields in forming galaxies and the
cumulative effect of XRBs on the ISM implies that they are likely
responsible for a non-negligible fraction of the field in mature
galaxies as well.

Along with the magnetization, it is clear that relativistic particles
from within the lobe will mix into the interstellar medium as cosmic
rays.  This process was discussed in more detail in \cite{heinz:02c}.
A critical question raised in that paper was the adiabatic cooling of
these particles.  The pressure inside the working surface has already
been estimated in eqs.~(\ref{eq:pdyn}) and the pressure inside the
trail/lobe to which the cosmic ray plasma has to expand is simply the
ISM pressure.

Taking the bulk of the particles leaving the working surface to have
energies comparable to the randomized specific kinetic energy of the
jet, i.e., $\gamma \sim \Gamma$, the final energy of any cosmic ray
protons in the trails is
\begin{equation}
  E_{\rm CR} \approx \gamma m_{\rm p} c^2 \left(\frac{p_{\rm ISM}}{p_{\rm
      dyn}}\right)^{\frac{1}{4}} \approx 3.3\,{\rm
    GeV}\,\cdot\,\Gamma_{5}\,\scrR^{\frac{1}{4}}
\end{equation}
indicating that adiabatic losses are likely small enough to leave the
results regarding the total energetic of cosmic ray injection by
microquasars discussed in \cite{heinz:02c} unchanged: the total
energetics of microquasars is likely to contribute of order a few
percent to the Galactic cosmic ray population.  Thus, microquasars
located in the vicinity of molecular clouds should be observable
through gamma-ray emission due to hadronic interaction with the cloud
protons detectable by GLAST.  A corollary of this process is GeV
neutrino emission that should also arise.  The high energy tail from
this emission might be detectable with high energy neutrino detectors
like Ice Cube.

\section{Observability}
\label{sec:observations}
Given the estimated pressure and size of the plasma trails and working
surfaces (i.e., hot spots), we can estimate their synchrotron
luminosity and surface brightness.  We once again use the expression
from eq.~(\ref{eq:bfield}) to parameterize the field strength.  Since
it is unclear what the composition of the plasma is ($e^{+}/e^{-}$
vs.~$p^{+}/e^{-}$), we parameterize the proton pressure as $p_{\rm
  p}=\xi_{\rm p}p_{\rm tot}/[(1+\xi_{B})(1+\xi_{\rm p})]$, where
$\xi_{\rm p}$ is the ratio of proton to lepton pressure.

\subsection{Radio trails}

While the field strength we estimate for the trails is not much larger
than the mean Galactic field, it is safe to assume that the partial
pressure from relativistic electrons inside the trails is larger than
outside, implying that the trails will produce synchrotron emission
that is significantly enhanced over the mean Galactic emission.

For a powerlaw distribution of electrons with index $2$ such that
$dN/d\gamma \propto \gamma^{-2}$, the synchrotron emissivity of the
plasma is \citep{rybicki:79}
\begin{equation}
  \epsilon_{\nu} \approx \frac{5 \times 10^{-38}\,{\rm ergs}}{\rm
  cm^{3}\,s\,Hz}\,\cdot\,\left(\frac{2
  p_{-11.5}}{1+\xi_{B}}\right)^{\frac{7}{4}} \frac{2
  \xi_{B}^{3/4}}{1+\xi_{\rm p}}\sqrt{\frac{\nu}{5\,{\rm GHz}}}
\end{equation}

The magnetic field strength inside the trail is approximately $B_{\rm
  eq} \approx 1.4 \times 10^{-5}\,{\rm mG}\,p_{-11.5}\,2\xi_{\rm
  B}/(1+\xi_{B})$, which gives a cooling time of
\begin{equation}
  \tau_{\rm cool} \sim 10^7\,{\rm yrs}\,\cdot\,\left[\frac{1 + \xi_{\rm
  B}}{2\xi_{\rm B} p_{-11.5}}\right]^{3/2}\nu_{5}^{-1/2}
\end{equation}
for an electron emitting at frequency $\nu \equiv 5\,{\rm
  GHz}\,\nu_{5}$.  The length of radio trail a particle can travel
before cooling losses affect it is
\begin{equation}
  l_{\rm cool} \sim \tau_{\rm cool}v_{\rm XRB} \approx 5\times
  10^{20}\,{\rm cm}\, \cdot\,\left[\frac{1 + \xi_{\rm B}}{2\xi_{\rm B}
  p_{-11.5}}\right]^{3/2}\frac{v_{17}\scrv}{\nu_{5}^{1/2}}
\end{equation}
which is comparable to $R_{\rm eq}$ from eq.~(\ref{eq:req}).  Thus,
the trails left behind by XRBs should typically only emit at
frequencies well below 5GHz.  Above such frequencies, the radio
emission from XRBs should be concentrated around the XRB itself.
Given their large angular sizes and low frequency emission, these
trails will be ideal targets for the new wave of high fidelity, low
frequency radio arrays, such as LOFAR, the LWA, the MWA, and,
ultimately, the SKA--low.

The central surface brightness of a given trail within a cooling
distance $l_{\rm cool}$ from the XRB will be
\begin{eqnarray}
  \Sigma & \approx & 2R_{\rm eq}\epsilon \\ & \approx & \frac{4\mu{\rm
      Jy}}{\rm arcsec^{2}}\left[\frac{2}{1+\xi_{\rm
      B}}\right]^{\frac{7}{4}}\frac{2p_{-11.5}^{\frac{5}{4}} \xi_{\rm
      B}^{\frac{3}{4}}}{1 + \xi_{\rm p}}
      \sqrt{\frac{W_{37.3}}{\nu_{5}v_{17}\scrv}} R_{M}^{-\frac{1}{8}}
      \nonumber
\end{eqnarray}
and the integrated trail luminosity at a fixed frequency is
\begin{equation}
  L_{\nu} \approx \pi\,R_{\rm eq}^2\,l_{\rm
    cool}\epsilon_{\nu}=3\times 10^{24}\,{\rm
    ergs\,Hz^{-1}\,s^{-1}}\,\cdot\,\nu_{5}^{-1}
\end{equation}

While every microquasar should have some kind of radio lobe/trail in
its vicinity, some sources will be better suited to study this effect
than others.  Cygnus X-1, for example, is already known to possess a
dark radio lobe.  As an HMXB with a relatively young age, it is
unlikely to show much of a radio trail, as argued above, and indeed,
the radio lobe seems to be close to the source, though it does show a
one-sided asymmetry: the radio ring surrounding it that led to the
claim of the radio lobe is one-sided.  This asymmetry could be due to
the fact the source is located at the border of a high density cloud.

Arguably one of the best sources to look for a radio trail is XTE
J1118+480, which is known to have both a steady low/hard state jet as
well as a large space velocity \citep{mirabel:01}.  We suggest that
this should be one of the first targets to study with low frequency,
large scale radio mapping for evidence of a plasma trail.

The large scale, asymmetric radio emission recently detected
surrounding Cygnus X-3 \citep{sanchez-sutil:08} is also suggestive and
we interpret it as dynamical interaction with the ISM.  The overall
morphology suggests that the source is moving in a roughly northern
direction relative to the ISM.  This should be followed up with VLBI
proper motion measurements of the XRB itself to test this hypothesis.

\subsection{Hot spots}
Given the estimate of the jet length $l_{\rm jet}$ from
eq.~(\ref{eq:length}) and the pressure inside the working surface, we
can estimate the order of magnitude of the synchrotron emission from
each hot spot as $L_{\nu,HS} \approx \epsilon_{\nu}\frac{4\pi(\alpha
  l)^3}{3}$.  Once the plasma has moved out of the working surface of
a jet, it expands (i.e., cools) adiabatically and reaches pressure
equilibrium with the lobe.  The pressure inside the hot spot is given
by
\begin{equation}
  p_{\rm HS} \approx \frac{W_{\rm jet}}{2\pi \left(\alpha\,l_{\rm
      jet}\right)^{2} c}
\end{equation}
giving an approximate the hot spot luminosoity of
\begin{eqnarray}
  L_{\rm HS,nu} & \approx & \epsilon_{\nu}\,\frac{4\pi (\alpha
  l)^3}{3} \\ & \approx & 4\times 10^{19}\,{\rm
  ergs\,s^{-1}\,Hz^{-1}}\frac{W_{37}^{3/2}p_{-11.5}^{1/4}\theta_{20}^{1/2}}
  {\alpha_{1}^{1/2}} \nonumber \\
  & & \ \ \ \ \ \ \ \ \cdot \left(\frac{28\Gamma\beta_{\rm jet}\scrR_{\rm
  M}}{25\Gamma + 15}\right)^{1/4} \nonumber
\end{eqnarray}

For typical microquasar parameters, the synchrotron cooling time
$\tau_{\rm X}$ for X-ray emitting electrons is longer than the
adiabatic cooling time and the same expression should hold for radio
as well as X-ray synchrotron emission.  This flux is accessible with
{\em Chandra} and in fact has been observed in a number of cases, most
notably XTE J1550 \citep{corbel:02}.  In the case of XTE J1550, the
hot spot has been observed to move away from the XRB at mildly
relativistic speeds, which fits in well with the picture developed
here: If the jet power rises sharply, as indicated by the radio flares
that precede the detection of the hot spots, the working surface will
(a) brighten and (b) move outward, decelerating as it moves into
undisturbed ISM.  The distances where the hot spots are observed are
consistent with where we would expect them from equation
(\ref{eq:length}).

\subsection{Bow shocks and shells}

As has been discussed in the literature \citep{gallo:05,russell:07},
and as is observed around radio galaxies in clusters
\citep[e.g.][]{birzan:04} the source will sweep up a quasi-spherical
shell during its initial inflation.  This shell will emit thermal
bremsstrahlung, potentially up to X-ray energies, as well as line
emission from the shocked plasma.  

For the early, quasi-spherical evolution and for stationary source, we
refer the reader to the discussion in the context of expanding shells
in galaxy clusters \citep[e.g.][]{heinz:98} and binaries
\citep{heinz:06d} and concentrate on the emission from the
ISM swept up ahead of radio trails in the case of XRBs with
appreciable space velocity.

Given that LMXB space velocities top out at a few hundred km/s, we
should expect temperatures up to about a keV, which should be
detectable by X-ray telescopes (which is also the case in pulsar-wind
nebulae).  The radiative cooling time strongly depends on the ISM
density, but assuming that the shock is fully radiative gives an upper
limit on the total luminosity from the swept up ISM of
\begin{eqnarray}
  L_{\rm shock} & < & \pi R_{\rm eq}^2 \frac{\rho_{\rm ISM}v_{\rm
  XRB}^3}{2} \\
  & \approx & 5\times 10^{35}\,{\rm ergs\,s^{-1}}\,\cdot\,v_{17}^2\,n_{1}\,
  \frac{W_{37}}{p_{-11.5}\scrv\scrR^{1/4}} \nonumber
\end{eqnarray}
spread over a region close to a degree in size (thus hard to detect
with the typical field of view of current X-ray instruments).

Longer wavelengths seem more suitable for finding these thermal
sources \citep[specifically, narrow band imaging of prominent
  shock-excited lines on degree scales, as well as broad-band imaging
  to look for free-free emission][]{russell:08}.  Given that the
estimated expansion velocity of the shell around Cyg X-1 is of order a
few dozen km/s, we would expect surface brightness and total power to
be comparable with those detected in the Cyg X-1 nebula.  This
suggests that asymmetric bow shock emission might be a suitable way to
identify sources for low frequency follow up.

\section{Conclusions}
\label{sec:summary}
We proposed that jets from X-ray binaries (microquasars) leave behind
trails of non-thermal, synchrotron emitting plasma as they move
through the interstellar medium.  LMXBs are more likely to leave such
trails due to their longer life expectancy and, more importantly, due to
the higher expected space velocities, while HMXBs likely produce more
stationary radio lobes.

The total plasma volume deposited by XRBs over the history of the
Galaxy is comparable to the total disk volume and constitutes a
non-negligible fraction of the halo volume.  We argue that a fraction
$f_{\rm mix}$ of order a few percent of this plasma is mixed into the
thermal ISM.

The magnetic field thus released can easily provide the seed field for
Galactic dynamos to produce the fields observed in spiral galaxies
even under the most conservative assumptions. Since HMXBs turn on
rapidly after an episode of star formation, this mechanism should be
important for galactic magnetic field production and maintenance
within a time frame of about $10^{7}\,{\rm yrs}$ from the first
supernovae.

We showed that radio emission from the plasma inside the trail should
be visible primarily at low frequencies, since the radiative cooling
time of the plasma at GHz frequencies limits emission to a roughly
spherical region around the binary.  LOFAR, MWA, LWA and other future
low frequency, wide field instruments are ideally suited for searches
of this emission.  LMXBs with high detected proper motions like XTE
J1118+480 will be the best candidates for such a search.

\acknowledgements{We would like to thank A. Merloni and C. Kaiser for
  helpful discussions. S.H. acknowledges support from NSF astrophysics
  grant 0707682.}

\end{document}